\documentclass[debug,overfull]{epl}
\usepackage{amssymb}

\newcommand{\Nav}{\mbox{$\langle N \rangle$}}

\shortauthor{A. Milchev \etal}
\shorttitle{Pressure in Giant Micelles}
\institute{
\inst{1}
Institute for Physical Chemistry, Bulgarian Academy of Sciences,
1113 Sofia, Bulgaria\\
\inst{2} D\'epartement de Physique des Mat\'eriaux,
Universit\'e Claude Bernard and CNRS, 69622 Villeurbanne Cedex, France\\
\inst{3}
Department of Applied Physics, Technische Universiteit Eindhoven,\\
Postbus 513, 5600 MB Eindhoven, The Netherlands\\
\inst{4}
Department of Physics \& Astronomy, University of Georgia,
Athens, Ga 30602, USA
}
\rec{}{}

\begin{document}

\title{Osmotic Pressure of Solutions Containing Flexible Polymers Subject to an
Annealed Molecular Weight Distribution}
\author{A. Milchev\inst{1} \and J.P. Wittmer\inst{2}
\thanks{E-mail: jwittmer@dpm.univ-lyon1.fr} 
\and P.~van der Schoot\inst{3} \and D.~Landau\inst{4}}

\maketitle

\begin{abstract}
The osmotic pressure $P$ in equilibrium polymers (EP) in good solvent is
investigated by means of a three dimensional off-lattice Monte Carlo
simulation. Our results compare well with real space renormalisation group
theory and the osmotic compressibility $K \propto \phi \upd \phi/\upd P$ from
recent light scattering study of systems of long worm-like micelles. We
confirm the scaling predictions for EP based on traditional physics of
quenched monodisperse polymers in the dilute and semidilute limit.
Specifically, we find $P\propto \phi^{2.3}$ and, hence, $K \propto
\phi^{-0.3}$ in the semidilute regime --- in agreement with both theory and
experiment. At higher concentrations where the semidilute blobs become too
small and hard-core interactions and packing effects become dominant, a much
stronger increase 
is evidenced and, consequently, the compressibility decreases much more
rapidly with $\phi$ than predicted from semidilute polymer theory, but again
in agreement with experiment.
\end{abstract}

\emph{Introduction.}
Superficially, it would seem that the colligative properties of quenched and
annealed polymers must differ vastly, because the former have a fixed molecular
weight distribution, whilst the latter are in equilibrium with each other, 
and continually exchange material \cite{florybook}. 
In this paper, where we study by means of computer simulations the osmotic 
pressure of ``equilibrium polymers" (EP) in a good solvent, 
we demonstrate that systems of EP, 
as widely varying as giant polymer-like surfactant micelles, and
supramolecular aggregates of dyes, dipolar colloids and proteins \cite{CC90}, 
behave in essence like conventional polymers.
The work presented here complements earlier simulation studies,
focusing on the density distribution of sizes in dilute,
semidilute and concentrated solution, covering both cases 
with \cite{WSMB00} and without ring closure \cite{WMC98,MWL00}. 
Our motivation for studying the osmotic pressure is that it is a physical
quantity that is more readily accessible experimentally than a distribution
function or a mean size, in particular in the regimes where the chains
strongly overlap. 
In addition, since conformation space renormalisation group (RG) 
predictions for the pressure are available \cite{PvS97}, as well
as experimental data of the osmotic compressibility covering at least two
concentration regimes \cite{Buhler}, a fairly comprehensive comparison
between theory, simulations and experiment is possible. The agreement we
find is remarkably good, strongly suggesting that our simulations not only
accurately describe the size distributions of EP, as in fact already shown
in previous work \cite{WMC98,MWL00,WSMB00}, but also capture their
colligative properties correctly.

After very briefly presenting the simulation method, we first present our
results for the pressure, and discuss these in the framework of the scaling
theory of polymers. 
Both systems with and without ring closure have been investigated. 
(We recall that in the magnetic analogy, used to
model EP, rings are suppressed \cite{degennesbook}.)
Our focus will mainly but not exclusively be on the findings
for the EP with suppressed ring closure. 
The reason is that ring closure is
thought not to play a significant role in 
giant micellar systems \cite{CC90,WSMB00},
for which the most extensive experimental data are available.
Also, for reasons to be discussed, although differences in the 
behavior of EP with
and without rings do show up in the dilute regime, the scaling functions
remain for all practical purposes the same. Also presented in this Letter is
the scaling behavior of the mean length of the EP, on the one hand because
it is directly related to the chemical potential of polymerizing materials
and on the other because it is subject of controversy \cite
{Schurtenberger}. The combined pressure and chemical potential data fully
describe the thermodynamics of the system in hand. We stress the success of
the ideas borrowed from the scaling theory of polymers in the dilute and
semidilute regimes, but also show that in the\ high-density limit the
properties of EP are dominated by packing effects, and not by chain
connectivity or self-assembly.

\begin{figure}[tbp]
\twofigures[scale=0.4]{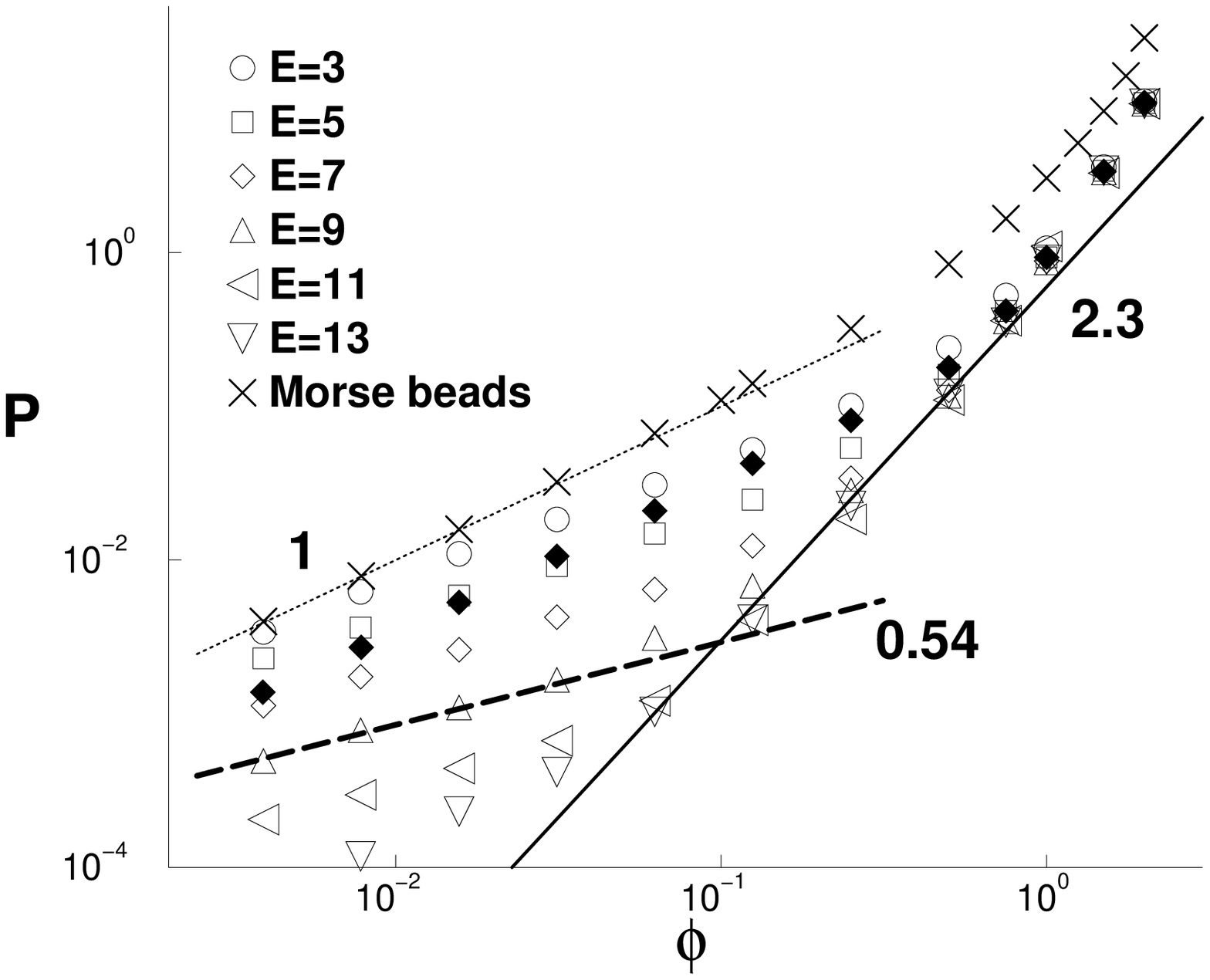}{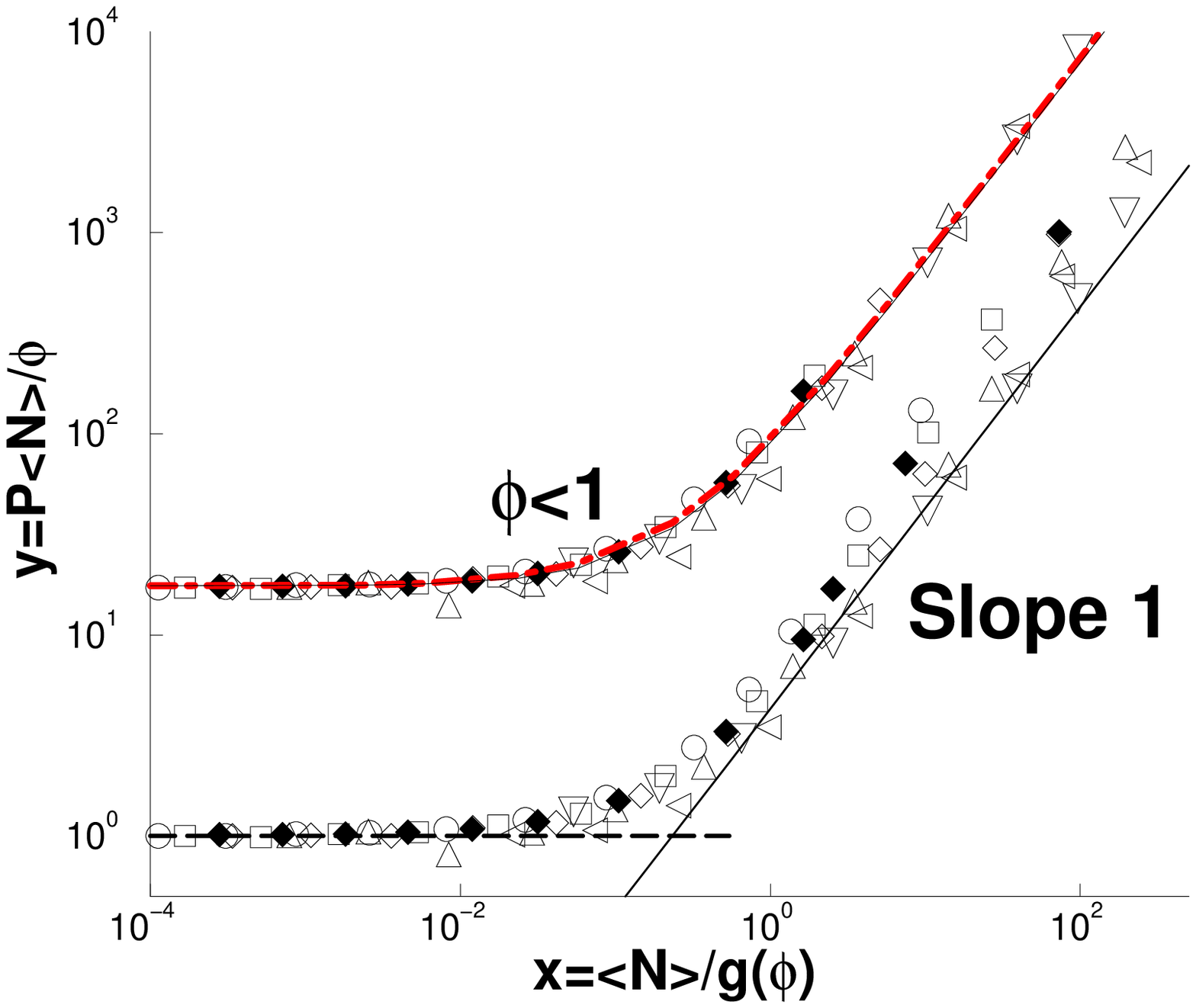}
\caption{Pressure $P$ \emph{versus} the number density $\phi $ for systems
of monodisperse Morse beads (crosses), equilibrium polymers with ring
closure suppressed (open symbols) and equilibrium polymers with ring closure
allowed (full symbols). Results for various scission energies $E$ are
indicated by the symbol shapes listed in the graph. The dashed and full
lines indicate the predicted scaling in the dilute and semidilute regimes
respectively, and the dotted line the ideal gas behavior of the Morse beads
at low densities. \label{fig:Pphi}}
\caption{
Master curve of the reduced pressure $y=P\left\langle
N\right\rangle /\phi $ \emph{versus} the reduced mean aggregation number $%
x=\left\langle N\right\rangle /g(\phi )$, based on the notions of the
scaling theory of polymers. Same symbols as in Fig.~\ref{fig:Pphi}. 
The bottom curve
gives all the measured data points, the top curve (shifted for clarity)
includes only those data for which $\phi <1$. The dashed line gives the
ideal gas law, and the bold line the power law, expected for a semidilute
solution. The data collapse is almost perfect when the data taken in the
high density regime (beyond the semidilute range) are excluded. In the top
curve we compare our simulations with the estimate $y=1+4.3x$ (thin line),
and the RG theory prediction from \protect\cite{PvS97} (dash-dotted line). 
Both curves virtually coincide.
\label{fig:Pscal}}
\end{figure}

\emph{Algorithm and some technical details.} As in our previous studies\cite
{MWL00,WSMB00}, we use an off-lattice Monte Carlo approach, generalizing an
efficient bead-spring model for polymers of fixed length.
For
the direct, non-bonding interactions between the beads we use the repulsive
part of a Morse potential. For all practical purposes one may view these
beads as hard spheres of diameter $d\approx 0.8$, and effective volume $%
v=\pi d^{3}/6\approx 0.25$. Here, and below, units of length are given in
units of the maximum bond length of the bonding potential between two
neighboring beads on a chain. This bonding interaction is described by the
so-called FENE (finitely extensible nonlinear elastic) potential. 
In addition to the FENE
potential we assign a scission energy $E>0$ to every bond, to be paid
whenever a bond is broken \cite{WMC98}. This scission energy is presumed to
be independent of the chain length and the density. In our presentation we
choose energy units such that the Boltzmann factor $k_{B}=1$, and work at
fixed temperature $k_{B}T=1$. 
The two operational
parameters which determine the system properties are the number density of beads 
$\phi $, and the scission energy $E$. Both have been varied over a wide
range, $E$ from $3$ to $13$, and $\phi $ from $2^{-8}$ to $2$
to produce a sufficient chain length and density
variation allowing for a critical test of the theoretical predictions \cite
{CC90,PvS97,WMC98}. We estimate that in our simulations the effective volume
fractions $v\phi $ probed range from $\approx 10^{-3}$ to $0.45$. The latter
value has to be compared with the (only slightly larger) hard-sphere
freezing volume fraction of about one half. Therefore, our simulations
extend all the way up to the melt regime of a dense liquid. The combined
FENE and Morse potentials fix the mean bond length at $l\approx d$,
essentially independent of chain length and density \cite{WSMB00}. In our
model the chains may be regarded as fully flexible \cite{WSMB00}. %

\emph{The osmotic pressure and its scaling with concentration.} %
%
As is usual in off-lattice Monte Carlo simulations, the pressure $P$ is
obtained by evaluating the virial, and adding to that the contribution
$k_B T \phi$ from the kinetic degrees of freedom \cite{MB94}. 
In Fig.~\ref{fig:Pphi} we
present the results for the bare pressure of the self-assembled polymers as
a function of the number density $\phi $, for various scission energies $E$.
For reasons of comparison results are also presented obtained for a gas of
monomeric beads, interacting solely via the Morse potential (``Morse
beads'').

For the Morse beads (indicated by crosses), we find ideal gas behavior at
densities $\phi \ll 1$. At about $v\phi \approx 0.1$ hard-core interactions
become important, and $P$ rises steeply. This is to be compared with our
results for EP without ring closure (open symbols), and with ring closure
(full symbols). Not surprisingly, switching on the polymerization in the EP
systems acts to lower the pressure relative to that of the gas of Morse
beads. Focusing on the EP without rings first, we observe that in the limit
of low densities and low scission energies, i.e., in the weak aggregation
limit, the pressure is comparable to that of the Morse beads, as it should
(see the data for $E=3$). As the scission energy and, hence, the mean chain
length increases, distinct power laws are observed in the dilute and
semidilute regimes.
The bold dashed line compares the pressure in the dilute limit with what one
expects on the basis of ideal gas behavior $P=\phi /%
\mbox{$\langle N
\rangle$}$, using the scaling of the mean chain length 
$\Nav \propto \phi ^{0.46}$ in the dilute regime derived below. At higher
densities the pressure becomes independent of $E$. In the limit of
sufficiently small densities, but large mean chain length, the measured
pressures converge towards $P \propto \phi ^{2.3}$. 
This is in line with the expected scaling behavior of
polymers (quenched or annealed) in semidilute solution. In semidilute
solution the pressure is governed by the number of blobs or entanglements
per unit volume, independent of molecular weight. It is therefore also
independent of any self-assembly process.

Of course, the blob picture is only valid for sufficiently low densities and
large chain lengths. Our simulation data confirm this: the larger the
scission energy $E$, the better the semidilute scaling regime is
discernible, in essence because the chains then overlap at lower densities.
At densities above $\phi \approx 1$, roughly when the packing effects
make themselves felt in the gas of Morse beads, the pressure of EP rises
more strongly than predicted by scaling theory, but still remains
independent of $E$. This is the melt regime discussed in the introduction.

%
In Fig.~\ref{fig:Pscal} we attempt to construct a master curve covering all
our data for the EP, essentially inspired by standard scaling theory for
conventional polymers \cite{degennesbook}, and elucidated in more detail in
the next section. The vertical axis gives the pressure divided by the
ideal gas pressure $\phi/\Nav$.
Given on the horizontal
axis is the ratio of the mean aggregation number 
\Nav \ and the number of monomers per blob 
$g \sim \phi ^{-1/(3\nu -1)}$,  
where $\nu
\approx 0.588$ is the Flory exponent of a self-avoiding walk in three
dimensions. 
%
%
The bottom curve
includes results from all densities probed, while the top curve excludes data
for densities $\phi \ge 1$, i.e., those data outside of the dilute and
semidilute regimes. Not surprisingly, the data collapse is only successful
if the data of the melt regime are excluded.

Included in both figures are also results obtained for EP in which
ring closure was allowed, and polymer rings compete with linear chains for
the available monomers. (The data are indicated by the filled symbols for a
single value of $E=7$.) In a previous study \cite{WSMB00} we found that in
the dilute limit most of the monomers are assembled in trimer rings (the
lower cutoff), and that the polymerization transition and the crossover to
the semidilute regime roughly coincide. The pressure data are consistent
with this, since we find at small densities $P=\phi /3$. Beyond the crossover
to the semidilute regime almost all of the additionally added monomers
are included in linear chains. The fact that a finite amount of monomers
remains bound in closed loops in the semidilute regime appears to be
irrelevant, for we observe within statistical accuracy the \emph{same}
pressure with and without rings. The data collapse perfectly in 
Fig.~\ref{fig:Pscal}, 
although a huge amount of trimer rings are present at low and
intermediate densities. Apparently, and perhaps surprisingly \cite{Cates98},
the presence of loops does
not significantly affect the excluded volume screening.

\emph{Theoretical considerations.} Our findings for EP without rings can be
understood from the following Ansatz for the grand potential density 
\begin{equation}
\Omega \lbrack c(N)]=\sum_{N=1}^{\infty }c(N)\left( \log (c(N)l^{3})-1-\mu
N+E+f_{N}(N,\phi )\right) -R(\phi ).  \label{eq:Omega}
\end{equation}
The sum over monomer mass $N$ describes the self-assembly in terms of an
ideal mixing entropy, a Lagrange multiplier or chemical potential $\mu
\leqslant 0$ that ensures conservation of mass, and a free energy penalty 
$E+f_N(N,\phi)$ associated with the chain ends \cite{WSMB00}. Of the
latter, $f_{N}(N,\phi )$ renormalizes the bare scission energy $E$ due to
differences in the micro-environment of the ends and central parts of the
chains. This term is different in the dilute, semidilute and concentrated
regimes \cite{WSMB00}. Applying the well-known statistical properties of
self-avoiding walks, we infer that in the dilute limit $f_{N}$ decreases
logarithmically with the degree of polymerization, i.e., 
$f_{N}(N)=-(\gamma -1)\log (N)$ with $\gamma \approx 1.158$ the
susceptibility exponent of the $n\rightarrow 0$ vector model in three
spatial dimensions \cite{degennesbook}. Entering the semidilute regime, one
expects excluded volume effects to be screened out, and $f_{N}$ should level
off to 
$f_{N}(\phi)=(\gamma-1)(3\nu-1)^{-1}\log(\phi)$ \cite{CC90}.
It is not known theoretically how $f_{N}(\phi )$ behaves in the melt
regime, so we focus first our analysis to the dilute and semidilute regimes.

The remaining contribution $R(\phi)$ accounts for excluded volume
interactions which depend only on density $\phi$, 
and are not conjugate to $c(N)$. 
Minimizing eq.(\ref{eq:Omega}) gives in the limit of long chains an exponential
equilibrium distribution function 
$c_{eq}(N)\propto \exp (-E-f_{N}(N,\phi)+\mu N)$ \cite{WSMB00}, 
and an osmotic pressure $P=-\Omega \lbrack
c_{eq}]=\phi /\mbox{$\langle N \rangle$}+R(\phi )$. 
Following De~Genes 
\cite{degennesbook}, we assume a constant free energy density of 
$k_{B}T \approx 1$ per blob, giving 
$P\mbox{$\langle N \rangle$}/\phi =F(x)=1+x$
with $x\propto \mbox{$\langle N \rangle$}/g(\phi )$. 
This simple estimate can fit remarkably well the observed master curve of 
Fig.~\ref{fig:Pscal},
and is barely distinguishable from the more elaborate 
RG treatment of ref.~\cite{PvS97}, also shown in the figure. 
Note that the
functional eq.~(\ref{eq:Omega}) may be readily generalized to include the
possibility of ring closure \cite{WSMB00}, or even chain branching. However,
as long as the excluded volume term $R(\phi)$ remains dominated by
entanglements, the equation of state should be independent of chain
architecture, in line with our observations.

\begin{figure}[tbp]
\twofigures[scale=0.4]{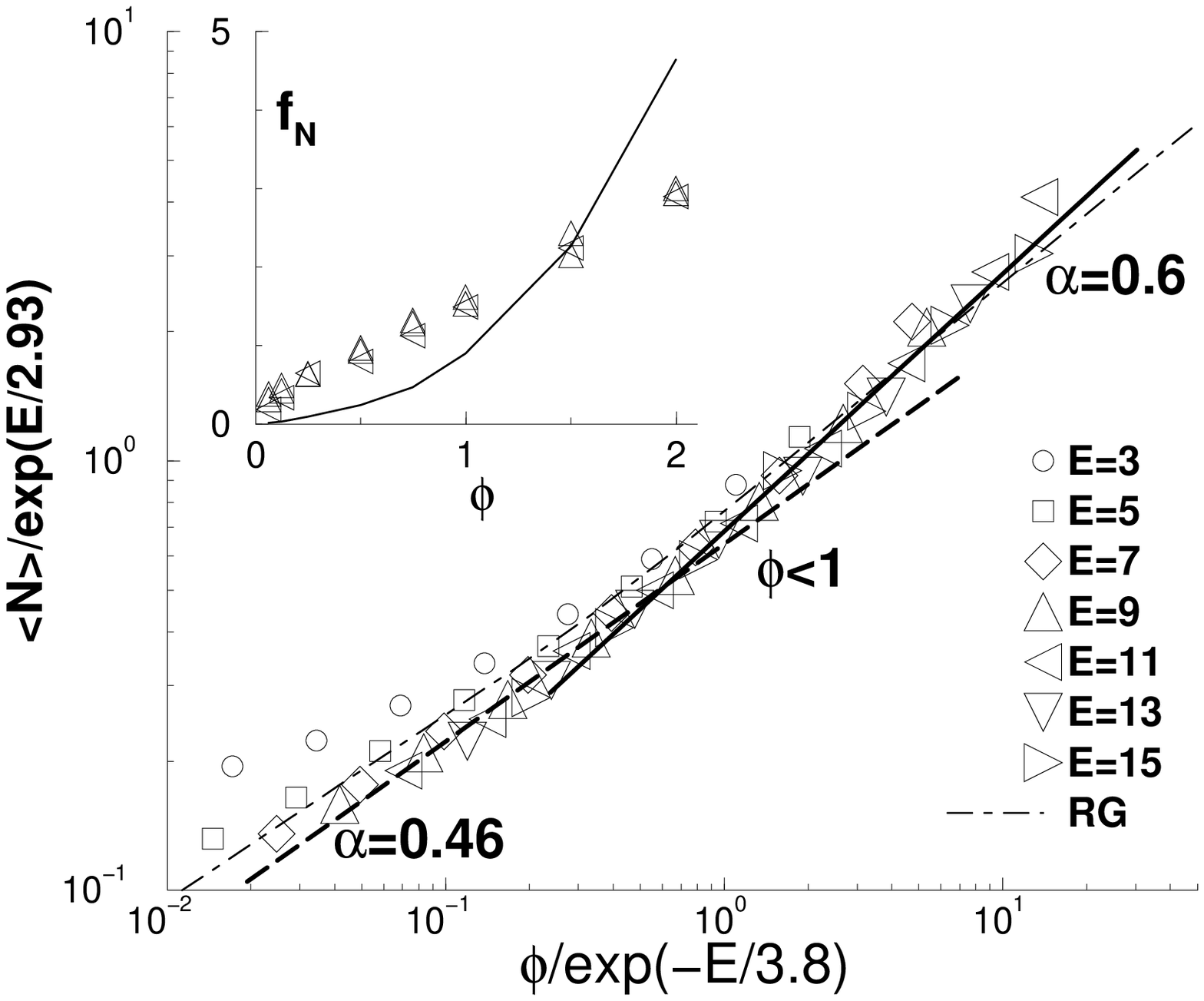}{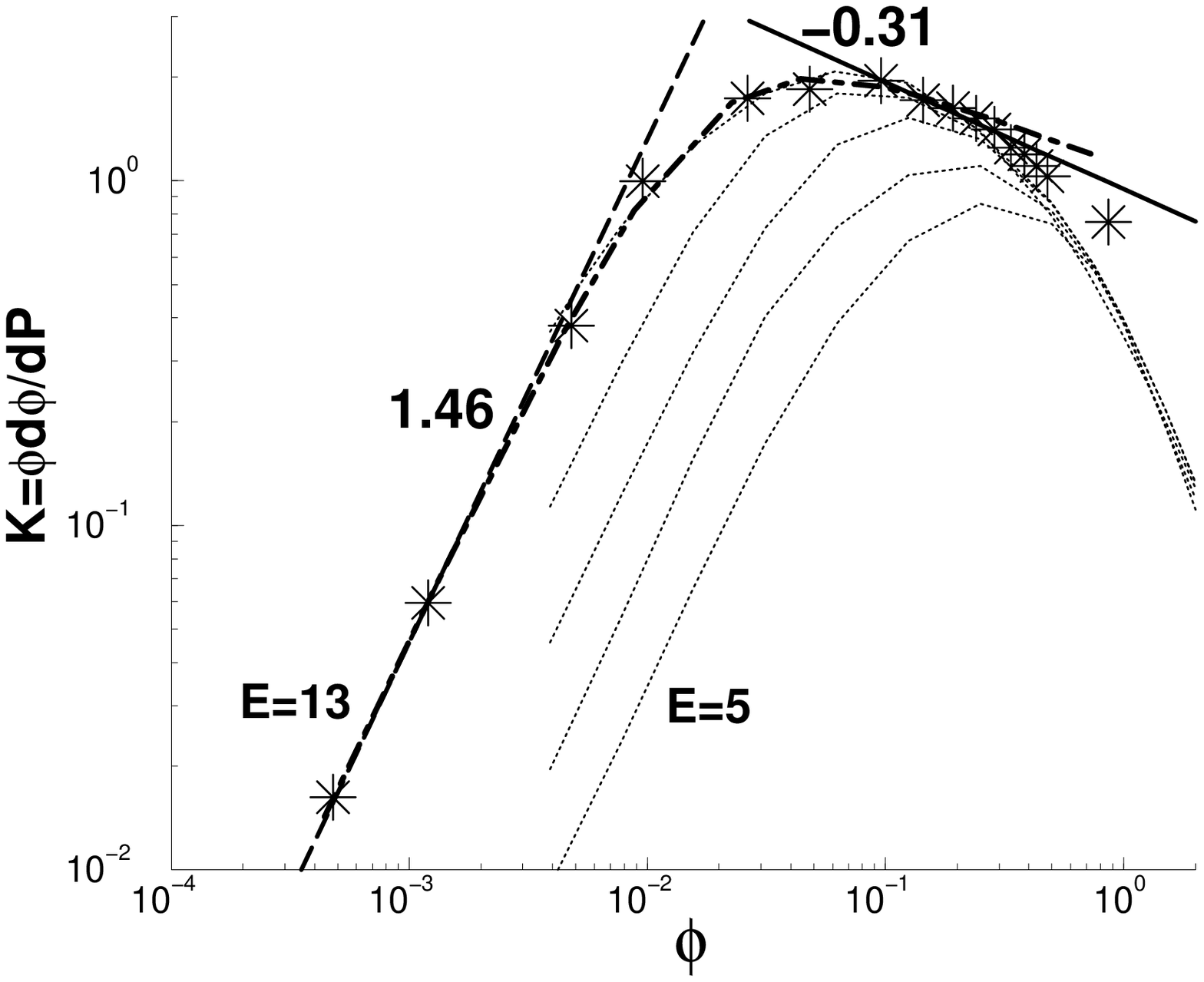}
\caption{Scaling of the reduced mean chain length 
$\Nav /\exp (E/2.93)$, \emph{versus} the reduced density $\phi /\exp (-E/3.8)$. 
Only data taken in the dilute and semidilute regimes are
shown ($\phi <1$). The dash-dotted line indicates the prediction from RG
theory \cite{PvS97}. Inset: concentration dependence of the free energy
penalty $f_N$
in the ``melt" regime, as obtained directly from the simulations
(triangles), and from the estimate $f_N=P/\phi $ (line).
\label{fig:Nscal}}
\caption{
Compressibility $K=\phi \partial P/\partial \phi $ \emph{versus}
the density $\phi $, from simulation (dotted lines), light scattering
measurements (asterisks) \cite{Buhler}, and from RG theory of EP
(dash-dotted line) \cite{PvS97}. 
The bold dashed line
gives the power law $K\propto \phi ^{1.46}$ in the dilute regime, the bold
line on the right the power law $K\propto \phi ^{-0.31}$ expected for a
semidilute solution. At higher densities, in the melt regime, the
compressibility drops much more rapidly due to packing restrictions.
\label{fig:Kphi}}
\end{figure}

\emph{Scaling behavior of mean chain length and chemical potential.} 
To help analyzing the
concentration and scission energy dependence of the pressure we evaluated
the mean chain length \mbox{$\langle N
\rangle$} as a function of these parameters, the results of which are given
in Fig.~\ref{fig:Nscal}. From the equilibrium length density distribution
given in the previous section, we easily find that 
$\Nav \propto \phi^{\alpha}\exp(\delta E)$ with 
$\alpha =\delta =1/(1+\gamma )\approx 0.46$ in the dilute, and 
$\alpha =(1+(\gamma-1)/(3\nu -1))/2\approx 0.6$, $\delta =1/2$ 
in the semidilute limit \cite{CC90,WMC98}.
This also fixes the chemical potential of the polymerizing material: 
$-1/\mu=\gamma \Nav$ in dilute, and $-1/\mu=\Nav$ in semidilute solution, 
at least for sufficiently large mean aggregation number. 
(The chemical potential has also been obtained directly from              
the exponential cutoff of the distribution functions, confirming            
these relations and yielding a similar scaling plot.)
In Fig.~\ref{fig:Nscal}, both the
mean aggregation number and the density are scaled by their values at the
crossover from the dilute to the semidilute regimes, obtainable by a
matching procedure described elsewhere \cite{WMC98}. 
The slopes giving the $\alpha$ exponents in both regimes are clearly visible. 
Also included in the figure is the RG result of \cite{PvS97}. 
As expected, the
agreement is reasonable but not perfect, because the theory is based on an
approximation valid in the vicinity of four spatial dimensions. Extrapolated
to three dimensions the growth exponents become $\alpha \approx 0.44$ in the
dilute, and $\alpha \approx 0.56$ in the semidilute regime, slightly
underestimating the results from our simulations in both regimes. The
astounding accuracy of the scaling function for the pressure calculated from
the RG theory, plotted in Fig.~\ref{fig:Pscal}, is due to the circumstance
that the inherent inaccuracies of the theory are hidden in the scaled
pressure and the scaled mean chain length.

As in Fig.~\ref{fig:Pscal}, we have in Fig.~\ref{fig:Nscal} only included data
for densities $\phi <1$ to avoid entering the melt regime, where the scaling
and RG theories are known to be no longer accurate \cite
{MWL00}. This is illustrated in the inset of Fig.~\ref{fig:Nscal}, where
we have plotted for three values of $E$ the concentration dependence of the
free energy penalty $f_{N}$, which in the simulations may be measured either
from the relation $\mbox{$\langle N \rangle$}^{2}\propto \phi \exp (E+f_{N})$
or from fitting the length distribution $c(N)$, 
which turns out to remain purely exponential in the melt regime
at variance with recent lattice simulations \cite{Yannick}
attempting to support a questionable modeling assumption 
proposed by Bouchaud \etal \cite{Bouchaud} to rationalize the 
observed Levy flight dynamical behavior in giant micelles.
The figure shows that in this regime
the free energy penalty $f_{N}$ increases more or less linearly with
concentration, $f_{N}(\phi )\propto \phi $. We speculate that packing
effects give rise to the free energy penalty, disfavoring chain ends
because they disturb the local structure of the chain fluid. A crude
estimate for $f_{N}$ can be gotten in terms of the work required to create
two new chain ends against the ambient pressure. 
In the inset of Fig.~\ref{fig:Nscal}, 
we compare $f_{N}$ with this estimate $P\delta V$, presuming
that the additional free volume $\delta V\approx 1/\phi $ associated with
free ends is of the order of the mean monomer volume. Although our argument
is certainly simplistic and further study is warranted, the qualitative
agreement does suggest that packing effects rule at such high densities --
this may of course also be inferred from Fig.~\ref{fig:Pphi}, 
which hints at the
convergence of the monomer and EP pressures in the melt regime.

\emph{Comparison with experiment: Osmotic compressibility.} 
In Fig.~\ref{fig:Kphi}
we compare our simulation results with the isothermal osmotic
compressibility, $K$, of a particular giant micellar system, measured by
means of light scattering \cite{Buhler}. 
%
We deduce $K$ from the computed pressures using the thermodynamic identity 
$K\propto \phi \partial \phi /\partial P$. In order to minimize scatter and
obtain a numerically meaningful differentiation, we first fit fourth order
polynomials to each data set of log $P$ \emph{versus} log $\phi $, and then
differentiate these polynomials. Because the dilute regime is very small for
large $E$, we have in addition extrapolated these data to lower densities
using the well tested relationship for the mean chain length in the dilute
limit (see left side of Fig.~\ref{fig:Nscal}). 
In other words, we insist on $K\propto \phi ^{1+\alpha}\propto \phi ^{1.46}$ 
in dilute solution for large $E$. 
The experimental data have been shifted in such a
way to match our semidilute regime where 
$K\propto \phi ^{-(2-3\nu )/(3\nu -1)}$. 
This is allowed, because 
(i) the experimental data for $K$ are not
absolute and given in arbitrary units only \cite{Buhler}, 
and (ii) our volume
fractions are only effective ones. Reasonable agreement was found over all
regimes for the scission energy $E=13$. From this we conclude that we are
able to simulate coarse-grained EP comparable in size to real giant
micelles. Also included is the result from the RG calculation \cite{PvS97}, 
which fits well to both experiment and simulation
in the dilute and semidilute regime. Apparently, the semidilute regime is
rather small in both simulations and in the experiments of Buhler and
co-workers \cite{Buhler}.

\emph{Discussion.} We have investigated the (osmotic) pressure and
compressibility of solutions of equilibrium polymers. Results from
off-lattice Monte Carlo simulation, renormalisation group theory, and recent
light scattering data of giant micelles have been compared. At low density
and sufficiently high scission energy, i.e., for long enough chains, the
results confirm unambiguously the theoretical predictions based on standard
polymer theory \cite{degennesbook}.

The universal scaling of $P$ and $K$ with respect to density $\phi$ and
mean chain-length \mbox{$\langle N \rangle$}\ is shown to be identical to
that of quenched polymers, in agreement with theory \cite{PvS97}.
Specifically, we find that in the semidilute regime the power law scaling
of the pressure $P\propto \phi ^{2.3}$ and compressibility $K\propto \phi
^{-0.3}$, in agreement with experimental findings in many solutions of giant
micelles \cite{CC90}, but only for sufficiently flexible ones
\cite{Berlepsch96,PvS95}.
Whether or not rings are present does not seem to alter the scaling of the
pressure in semidilute EP. 
Despite that there is consensus about the growth law in the dilute regime, 
now and again workers find growth exponents that deviate from the expected 
value close to one-half, varying from $\alpha \approx 0.1$ to $1.5$
\cite{PvS95,Schurtenberger,Jerke99,Stradner00}. 
From our simulations we put forward that for these systems
additional physics must be involved, such as due to ring
closure, branching, the presence of charges, or the presence of more than
one aggregating component. Clearly, the issue calls for a more detailed
experimental investigation.

In our simulations the blob picture breaks down at volume fractions
exceeding, say, $0.1$. The pressure then increases more rapidly than with
the $2.3$ power law, although it does stay below that of the monomeric Morse
beads at equal concentration. As a consequence, the deduced compressibility
decreases more strongly in this limit than with the $-0.3$ power expected
from scaling theory. Our interpretation is that the EP then enter the
concentrated or melt regime. The light scattering data of Buhler \etal \ 
also
point at a crossover to the concentrated regime \cite{Buhler}. In real
systems the crossover to the concentrated regime is non-universal, and
likely to depend on both the persistence length of the aggregates as well as
their diameter \cite{GK94}.
We intend to investigate by means
of computer simulation the influence of chains flexibility on the
colligative properties of EP in the near future.

\acknowledgments
We thank M.E.~Cates and J.-L.~Barrat for stimulating discussions.


\end{document}